\documentclass{ifacconf}
\usepackage{balance}
\usepackage{graphicx}
\newif\ifusebiblatex
\usebiblatexfalse      
\ifusebiblatex
\usepackage[
    natbib=true,
    maxbibnames=3,
]{biblatex}
\else
\usepackage{natbib}        
\fi
\usepackage{xcolor}
\usepackage{mymath} 

\usepackage{mathrsfs}
\usepackage{parskip} 
\usepackage{aligned-overset}
\usepackage{nicematrix}
\usepackage[noadjust]{cite}
\usepackage{siunitx}
\usepackage{booktabs}
\usepackage{multirow}
\usepackage{tikz}

\usepackage{graphicx}
\usepackage{subcaption}
\usepackage{varwidth, tikz, nth}
\usepackage{pgfplots}
\usepgfplotslibrary{patchplots,colormaps,colorbrewer,fillbetween}
\pgfplotsset{compat=newest}
\usetikzlibrary{shapes, arrows, shapes.misc, arrows.meta, positioning, matrix, calc, fit, fadings, patterns, pgfplots.colorbrewer,decorations.markings,decorations.pathmorphing,backgrounds,intersections}
\usetikzlibrary{shapes.geometric}
\usetikzlibrary{decorations.text}
\usepgflibrary{decorations.markings} 
\definecolor{legreddraw}{HTML}{E41A1C}
\definecolor{legbluedraw}{HTML}{377EB8}
\definecolor{leggreendraw}{HTML}{4DAF4A}
\definecolor{legvioletdraw}{HTML}{984EA3}
\definecolor{legredfill}{HTML}{B61516}
\definecolor{legbluefill}{HTML}{2C6593}
\definecolor{leggreenfill}{HTML}{3E8C3B}
\definecolor{legvioletfill}{HTML}{7A3E82}
\usepackage{xurl}

\newif\ifarxiv
\arxivtrue      
\ifarxiv
\newcommand{\appendixcite}[2]{Appendix~\ref{#1}}
\newcommand{\appendixeqref}[2]{\eqref{#1}}
\else
\newcommand{\appendixcite}[2]{\citep[][Appendix~#2]{perreault2026arxiv}}
\newcommand{\appendixeqref}[2]{\citep[][Eq.~(#2)]{perreault2026arxiv}}
\fi

\begin{document}
\begin{frontmatter}

\title{Host-Aware Control of Gene Expression using Data-Enabled Predictive Control} 

\thanks[footnoteinfo]{This work was supported by EPSRC under the EEBio Programme Grant, EP/Y014073/1.}

\author[]{Liam Perreault, Idris Kempf$^{\dagger}$, Kirill Sechkar,} 
\author[First]{Jean-Baptiste Lugagne \& Antonis Papachristodoulou} 

\address[First]{University of Oxford, Parks Road, Oxford OX1 3PJ, UK ($\,^{\dagger}\!$corresponding author: idris.kempf@eng.ox.ac.uk).}

\begin{abstract}
Cybergenetic gene expression control in bacteria enables applications in engineering biology, drug development, and biomanufacturing. AI-based controllers offer new possibilities for real-time, single-cell-level regulation but typically require large datasets and re-training for new systems. Data-enabled Predictive Control (DeePC) offers better sample efficiency without prior modelling. We apply DeePC to a system with two inputs---optogenetic control and media concentration---and two outputs---expression of gene of interest and host growth rate. Using basis functions to address nonlinearities, we demonstrate that DeePC remains robust to parameter variations and performs among the best control strategies while using the least data.
\end{abstract}


\end{frontmatter}

\makeatletter
\let\old@endfrontmatter\endfrontmatter
\renewcommand{\endfrontmatter}{%
  \old@endfrontmatter
  \endNoHyper
}
\makeatother

\section{INTRODUCTION}

Cybergenetics combines engineered biological circuits with computer-based controllers, enabling precise control of cellular processes for fundamental research, biomanufacturing, or the development of new biological circuits~\citep{KHAMMASHREVIEW}. The controlled quantities usually comprise gene expression levels, often measured indirectly via linked fluorescent reporter proteins, but they can also include other cellular variables, such as cell growth rate or metabolite production. These processes are externally actuated using optogenetic, chemical, or mechanical inputs. Due to the inherent complexity of biological processes, current state-of-the-art methods for cybergenetics either rely on model-based controllers~\citep{automated_optogenetic_control_milias_argeitis_206,shaping_bacterial_populationChait2017}, which are constrained by the validity of (linearised) models, or AI-based methods~\citep{deep_mpc_thousands_of_single_cells_Lugagne2024,10178155,ESPINELRIOS202561,ESPINELRIOS2026}, which are labour-intensive to train and must be re-trained for each application.

In this paper, we leverage recent developments in data-enabled predictive control (DeePC)~\citep{DEEPC} to develop a plug-and-play cybergenetic controller for engineering biology applications. Our approach scales to the control of thousands of cells in parallel without requiring system identification or prior training. Although we base our algorithm on \textit{linear} DeePC for computational efficiency, we introduce basis functions to capture the sigmoidal nonlinearities commonly encountered in biological systems~\citep{10591192}. We demonstrate that the controller is robust to parameter uncertainty in the sigmoidal function and measurement noise. Furthermore, we explore improving data efficiency by using model reduction techniques. Finally, we provide a comprehensive comparison with model-free (proportional-integral control), model-based (successive linearisation model predictive control~\citep{slmpc_zhakatayev2017}), AI-based (deep model predictive control~\citep{deep_mpc_thousands_of_single_cells_Lugagne2024}) control strategies, and reinforcement learning~\citep{lillicrap2019continuouscontroldeepreinforcement}. 

The controller is applied to a two-input two-output biological system in \textit{E. coli} that incorporates the light-sensitive \textit{CcaS/CcaR} system~\citep{Olson2014}---a genetic toggle switch that is used to activate the expression of genes of interest. The controlled quantities are assumed to be synthetic gene expression, measured by gauging the emission of a fluorescent reporter protein, and the cell's growth rate. Unlike prior research that introduces additional synthetic circuits~\citep{automated_optogenetic_control_milias_argeitis_206,Barajas2022} or uses external dilution~\citep{pmlr-v242-brancato24a} to control the growth rate, we assume that the growth rate is controlled through the medium's nutrient density. To capture our system's dynamics, we therefore develop a host-aware model that integrates a coarse-grained bacterial host model~\citep{weisse_mechanistic_links} with a mechanistic model of the \textit{CcaS/CcaR} system and downstream genes of interest. Although the resulting model comprises 18 states and includes nonlinearities, we show that DeePC can successfully control both gene expression and growth rate with high sample efficiency. These results inform the design of data-driven cybergenetic controllers, providing guidelines for optimal controller hardware and experimental design.

The paper is organised as follows. In Section~\ref{sec:modelingsystemproperties}, we outline our host-aware biological process model of the system being modelled. In Section~\ref{sec:deepcsection}, we design a DeePC algorithm using basis functions and assess its robustness against noise and parameter uncertainty in Section~\ref{sec:robustness}. Finally, Section~\ref{sec:comparison} compares the DeePC algorithm with other control methods in terms of performance, as well as sample and computational efficiency. 

\paragraph*{Notation} For scalars, vectors or matrices $A_i$, let $A_i\otimes A_j$ denote the Kronecker product and $\col(A_1,\dots,A_n)=[A_1^\Tr,\dots,A_n^\Tr]^\Tr$ their vertical concatenation. For a vector $a\in\R^p$, let $\norm{a}_Q\eqdef \sqrt{\trans{a}Q{a}}$, $\inR{Q}{p}{p}$, denote the weighted 2-norm and $\onenorm{a}\eqdef \sum_{i=1}^p \abs{a_i}$ the 1-norm.

\section{MODELLING AND SYSTEM PROPERTIES}\label{sec:modelingsystemproperties}


To analyse data-based controllers for biological systems, we develop a model that captures the coupling between cell growth and synthetic gene expression under external inputs. We use a published coarse-grained \textit{E.~coli} cell model \citep{weisse_mechanistic_links,nikolados_host_circuit_models} to simulate bacteria hosting a synthetic GFP reporter gene, which is optogenetically regulated by the \textit{CcaS/CcaR} system~\citep{Olson2014}. The nonlinear model capturing the expression of the set $\mathcal{S}$ of all considered genes and the cell's energy metabolism can be found in \appendixcite{sec:app:model}{A}. 

The first input is the externally provided nutrient's concentration $u_\mathrm{s}$. Internalised nutrients are converted into energy storage molecules $a$ necessary for all gene expression processes in the cell{\ifarxiv\footnote{For any variable $x$, its name is used to refer to both biochemical species $x$ and its respective cellular concentration.}\fi}. The first output is the rate of cell growth $y_\lambda = \lambda$, which according to \eqref{eqn:growth} is related by the cell's protein density $\rho$ to
the total rate of all mRNAs' translation $\{M_\mathrm{x}\}_{\mathrm{x}\in\mathcal{S}}$. Here, where $M_\mathrm{x}$ denotes the concentration of translational complexes and $\gamma(a)$ is an energy-dependent elongation rate (see \appendixcite{sec:app:model}{A}).
\begin{equation}\label{eqn:growth}
\lambda(a,\{M_\mathrm{x}\}_{\mathrm{x}\in\mathcal{S}})\eqdef 
\left(\gamma(a)/\rho\right)\sum_{\mathrm{x}\in \mathcal{S}} M_\mathrm{x},
\end{equation}

The input $u_g$ is the ratio of red and green light intensities shone upon the cell, affecting the production rate of GFP mRNA $\alpha_\text{syn}$ as per \eqref{eqn:synthetic_transcription}. Here, $\alpha_{\text{syn},\text{max}}$ is the maximum transcription rate, $\theta_\text{syn}$ and $A_g$ are the energy and input half-saturation constants, $F_\text{b}$ is the baseline rate of transcription, and $\tau_\mathrm{g} \geq 0$ is a time delay accounting for the unmodelled \textit{CcaS/CcaR} dynamics~\citep{rullan2018optogenetic}: 
\begin{align}
\alpha_\text{syn}(a,u_\mathrm{g})\eqdef
    \frac{\alpha_{\text{syn},\text{max}}a}{\theta_\text{syn}+a} \cdot 
    \frac{F_b +\left(u_\mathrm{g}(t-\tau_\mathrm{g})\right)^{h_\mathrm{g}}}{A_\mathrm{g}+\left(u_\mathrm{g}(t-\tau_\mathrm{g})\right)^{h_\mathrm{g}}}.\label{eqn:synthetic_transcription}
\end{align}
Transcribed GFP mRNAs $M_\mathrm{g}$ are translated at the energy-dependent rate $v_\mathrm{g}(a)$ into nascent proteins $p_\mathrm{g}$, which maturate at the rate $\mu_\mathrm{g}$ as per \eqref{eqn:synthetic_proteins}. The level of mature (i.e. fluorescent) GFP $y_\mathrm{g}=P_\mathrm{g}$ represents the second output. Importantly, synthetic GFP expression both consumes and depends on the energy $a$ obtained from nutrients $u_\mathrm{s}$. Moreover, all species' concentrations are affected by dilution as the cell grows in volume at rate $\lambda$, which in turn is related to $M_\mathrm{g}$ according to \eqref{eqn:growth}:
\begin{align}\label{eqn:synthetic_proteins}
&\dot{p}_\mathrm{g} = v_\mathrm{g}(a) M_\mathrm{g} - (\lambda + \mu_\mathrm{g}) p_\mathrm{g},
&\dot{P_\mathrm{g}} = \mu_\mathrm{g} p_\mathrm{g} - \lambda P_\mathrm{g},.
\end{align}
The overall dynamics can be summarised as:
\begin{align}\label{eq:nonlinearsys}
&\dot{x}=f(x,u),
&y=h(x),
\end{align}
where $x\in\!\R^{n_x}$ with $n_x=18$ is the column state vector for the system, $u\!\eqdef\!\col(u_s/\bar{u}_s, u_\mathrm{g}/\bar{u}_\mathrm{g})\!\in\!\R^{n_u}$, $n_u=2$ and $y\!\eqdef\!\col(y_{\lambda}/\bar{y}_\lambda, y_{\mathrm{g}}/\bar{y}_{\mathrm{g}})\!\in\!\R^{n_y}$, $n_y=2$ (see \appendixcite{sec:app:model}{A} for normalisation factors $\bar{u}$ and $\bar{y}$). The aim of control is to find inputs $u$ to track time-varying references $y_{\lambda,\text{ref}}$ and $y_{\mathrm{g},\text{ref}}$.

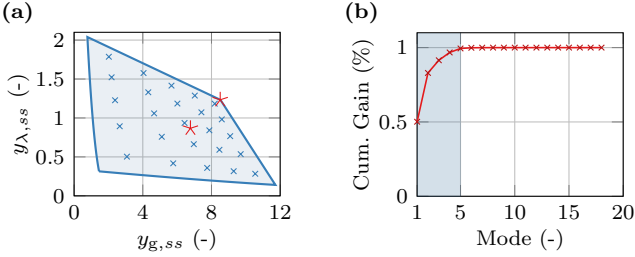
\begin{figure}
\centering
\begin{subfigure}[b]{0.49\linewidth}
\centering
\begin{tikzpicture}
\begin{axis}[axis on top=false,clip=false,font=\footnotesize,width=0.99\linewidth,height=3.75cm,
xlabel={$y_{\mathrm{g},ss}$ (-)},
ylabel={$y_{\lambda,ss}$ (-)},
grid=both,
mark size=3pt,
xmin=0,xmax=12,ymin=0,ymax=2.1,
xlabel style={yshift=0.3em}, xtick={0,4,8,12},
ylabel style={yshift=-0.5em},
]
\addplot[thick,color=legbluedraw,
  fill=legbluefill,   
  fill opacity=0.1] 
table[x=x_norm,y=y_norm,col sep=comma]
{csv/reachable_set_hull.csv};

\addplot[only marks,mark=x,color=legbluedraw,mark size=1.5pt] 
table[x=x_ref_norm,y=y_ref_norm,col sep=comma] {csv/ref_data_norm.csv};

\addplot[only marks,mark=star,mark size=3pt,color=legreddraw] 
coordinates {(6.773520263313702,0.867595844882394) (8.507312905353169,1.233582969288464)};

  \node[anchor=south west,font=\footnotesize]
    at (rel axis cs:-0.4,1) {\textbf{(a)}};
\end{axis}

\end{tikzpicture}
\label{fig:sub1}
\end{subfigure}
\hfill
\begin{subfigure}[b]{0.49\linewidth}
\centering
\begin{tikzpicture}
\begin{axis}[axis on top=false,clip=false,font=\footnotesize,width=0.99\linewidth,height=3.75cm,
xlabel={Mode (-)},
ylabel={Cum. Gain (\si{\percent})},
grid=both,
mark size=3pt,
xmin=1,xmax=20,ymin=0,ymax=1.1,xtick={1,5,10,15,20},
xlabel style={yshift=0.3em},  
ylabel style={yshift=-0.5em},
]
\addplot[on layer=axis background,draw=none, fill=legbluefill, fill opacity=0.2] coordinates {(5,0) (5,1.1) (1,1.1) (1,0) (5,0)};
\addplot[semithick,color=legreddraw] 
table[x=number_of_modes,y=minimum_percent_dynamics,col sep=comma]
{csv/balanced_reduction_grid.csv};
\addplot[draw=none, mark=x, mark size=1.5pt,color=legredfill] 
table[x=number_of_modes,y=minimum_percent_dynamics,col sep=comma]
{csv/balanced_reduction_grid.csv};

  \node[anchor=south west,font=\footnotesize]
    at (rel axis cs:-0.4,1) {\textbf{(b)}};
\end{axis}
\end{tikzpicture}
\label{fig:sub2}
\end{subfigure}\\[-1em]
\caption{System properties: (a) steady-state outputs for constant inputs and (b) cumulative gains for modes of the model reduction. Crosses and circles in (a) denote points used for performance evaluation, respectively.}\label{fig:fig1}
\end{figure}

To determine the range of reachable outputs, \eqref{eq:nonlinearsys} is simulated in open loop for a 2D grid of constant inputs in the ranges $u_s/\bar{u}_s\in[10^{-2},5]$ and $u_g/\bar{u}_\mathrm{g}\in[0,4]$, with parameter values taken from the literature~\citep{weisse_mechanistic_links, nikolados_host_circuit_models, refactoring_light_switchable_tcs_Schmidl2014,bionumbers_Milo2010}. The resulting steady-state outputs $y_{\mathrm{g},ss}$ and $y_{\lambda,ss}$ are shown in Fig.~\ref{fig:fig1}.a. For larger growth rates $y_{\lambda,ss}$, the circuit's output $y_{\mathrm{g},ss}$ decreases due to the coupling through dilution and shared resources~\citep{weisse_mechanistic_links}.

The nonlinear system~\eqref{eq:nonlinearsys} can be linearised at arbitrary points $x_\star$ and $u_\star$, yielding the state-space representation
\begin{align}\label{eq:linearsys}
&\dot{x}=\bar{A} x + \bar{B} u + \bar{f}_\star,
&y=\bar{C}x + \bar{h}_\star.
\end{align} 
where $\bar{A}_{i,j} \!=\! \partial f_i/\partial x_j |_{x_\star,u_\star}$, $\bar{B}_{i,j}\!=\!\partial f_i/\partial u_j|_{x_\star,u_\star}$, $\bar{C}_{i,j} \!=\!\partial h_i/\partial x_j|_{x_\star}$, $\bar{f}_\star=f(x_\star,u_\star) - Ax_\star -Bu_\star$, and $\bar{h}_\star = h(x_\star,u_\star) - Cx_\star$. Model~\eqref{eq:linearsys} has $n_x=18$ states, many of which are not observable for the linearisation points. For the subsequent controller synthesis, a balanced model reduction is therefore performed~\citep[Ch. 9.2]{Brunton_Kutz_2022}. First, we compute the balanced Gramians for~\eqref{eq:linearsys} at the steady states from Fig.~\ref{fig:fig1}.a. We then compute the minimum cumulative sum of Gramian eigenvalues (across all steady states) of the $n_x=18$ balanced modes, which are shown in Fig.~\ref{fig:fig1}.b. This plot shows that with only $5$ states, \SI{99.4}{\percent} of the most observable and controllable portion of the linear dynamics are captured, which can be exploited to reduce the amount of training data for control (see Section~\ref{sec:amount_of_data}). Note that the reduced system capture the most observable and controllable dynamics, but its states lose their biochemical meaning.

In the following, it will be assumed that the inputs are constant for $t\in[kT_s,\,(k+1)T_s)$, where $T_s=\SI{10}{\minute}$ is the sample time, and an equivalent discrete-time representation of~\eqref{eq:linearsys} will be used:
\begin{align}\label{eq:linearsysdt}
&x_{k+1}=A x_k + B u_k + f_\star,
&y_{k+1}=C x_{k+1} + h_\star.
\end{align} 

\section{Data-Enabled Predictive Control}\label{sec:deepcsection}


Model predictive control (MPC) has already been applied to (single-input) gene regulation problems considered here (see e.g.~\citep{automated_optogenetic_control_milias_argeitis_206} or~\citep{deep_mpc_thousands_of_single_cells_Lugagne2024}). Given system~\eqref{eq:linearsysdt}, an MPC scheme computes an input sequence by predicting the future evolution of the system and optimising an objective function over some planning horizon $N$~\citep{maciejowski2002predictive}. The first input of this sequence is applied to the system and the optimisation repeated at time $t+1$. Although the advantages of MPC are widely recognised, standard schemes rely on an accurate model to predict the future evolution of a system. In contrast, here we consider data-enabled predictive control (DeePC) that bypasses modelling by predicting the future evolution of the system from measured input-output data~\citep{DEEPC,8933093}. Let $\bm{u}^d\!\eqdef\!\col(u_1^d,\dots,u_{T}^d)$ be a sequence of inputs applied to the dynamical system, $\bm{y}^d\!\eqdef\!\col(y_1^d,\dots,y_{T}^d)$ the corresponding sequence of measured outputs, and define
\begin{equation}\label{eq:data}
\begin{pmatrix} U_{\text{p}}\\ U_{\text{f}} \end{pmatrix} \eqdef \mathscr{H}_{T_{\text{ini}}+N}(\bm{u}^{\text{d}}),
\quad\begin{pmatrix} Y_{\text{p}}\\ Y_{\text{f}} \end{pmatrix}\eqdef\mathscr{H}_{T_{\text{ini}}+N}(\bm{y}^{\text{d}}),
\end{equation}
where $U_{\text{p}}$ and $U_{\text{f}}$ consists of the first $T_{\text{ini}}n_u$ and last $Nn_u$ rows of the Hankel matrix $\mathscr{H}_{T_{\text{ini}}+N}(\bm{u}^{\text{d}})$ (and similarly for $Y_{\text{p}}$ and $Y_{\text{f}}$). 

Suppose that $T \geq (n_u+1)(T_\text{ini}+N+n_x)-1$ and $T_{\text{ini}}\geq\ell$, where $\ell$ is the lag of the system, and the input $\bm{u}^d$ is persistently exciting of order $T_{ini}+N+n_x$ (\appendixcite{sec:app:hankel}{B}), then a regularised DeePC scheme solves the following data-based optimisation problem:
\begin{equation}\label{eq:deepc}
\begin{aligned}
\min_{g, \bm{u}, \bm{y},\sigma_{y}}\,& \norm{\bm{y}-\bm{r}}_{Q}^{2}+\norm{\bm{u}}_{R}^{2} +\rho_{g}\norm{g}_{1}+\rho_{y}\norm{\sigma_{y}}_{1}\\ 
\text{s.t.}\,&\begin{pmatrix} U_{\text{p}}\\ Y_{\text{p}}\\ U_{\text{f}}\\ Y_{\text{f}} \end{pmatrix}g\!=\!
\begin{pmatrix} \bm{u}_{\text{ini}}\\ \bm{y}_{\text{ini}}\\ \bm{u}\\ \bm{y} \end{pmatrix}\!+\!\begin{pmatrix} 0\\ \sigma_{y}\\ 0\\ 0 \end{pmatrix}\!,\quad \bm{u}\!\in \!\mathcal{U}_N,
\end{aligned}
\end{equation}
where $\bm{u}\!\eqdef\!\col(u_t,\dots,u_{t+N-1})$, $\bm{y}\!\eqdef\!\col(y_t,\dots,y_{t+N-1})$, and $\bm{r}\!\eqdef\!\col(r_{t},\dots,r_{t+N-1})$ are the input, output, and reference signals, $\bm{u}_{\text{ini}}\!\eqdef\!\col(u_{t-T_\text{ini}},\dots,u_{t-1})$ and $\bm{y}_{\text{ini}}\!\eqdef\!\col(y_{t-T_\text{ini}},\dots,y_{t-1})$ initial trajectories, $\sigma_{y}\!\in\!\R^{T_{ini} n_y}$ slack variables, and $Q$, $R$, $\rho_g$, and $\rho_y$ fixed weights of appropriate dimensions. For the following application, output constraints are omitted and the input constraint set is defined as $\mathcal{U}_N \eqdef \left\lbrace \bm{u}\in\R^{N n_u} \mid u_\text{min} \leq u_k \leq u_\text{max},\right\rbrace$, where $u_\text{min}\!\eqdef\!\col(10^{-2},0)$, $u_\text{max}\!\eqdef\!\col(5,4)$, and $k=0,\dots,N-1$. Like for standard model-based MPC, the DeePC algorithm solves~\eqref{eq:deepc} in a receding horizon manner, resulting in feedback control. 

Instead of penalising the inputs, the cost function in~\eqref{eq:deepc} is modified to penalise the input difference in two subsequent timesteps. We rewrite $\bm{u}\!\eqdef\!\bm{u_{t-1}} + \Delta \bm{\delta u}$, where $\bm{u_{t-1}}\!\eqdef\!\col(u_{t-1},\dots, u_{t-1})$, $\bm{\delta u}\!\eqdef\!\col(\delta u_{t}, \dots, \delta u_{t+N-1})$ is the vector of input changes, and $\inR{\Delta}{n_u}{n_u}$ is block lower-triangular with blocks $I_{n_u}$. Problem~\eqref{eq:deepc} then optimises over $\delta \bm{u}$ instead of $\bm{u}$ and $\mathcal{U}_N$ is updated accordingly at each time step. In the absence of other input constraints and penalties, this has the effect of introducing integral action~\citep[Ch. 2.4]{maciejowski2002predictive}. The weight matrices and scalars in~\eqref{eq:deepc} are tuned using simulations and unless otherwise noted, chosen as $Q=I_N\otimes\diag(10^{-1},1)$, $R=I_N\otimes\diag(10^{-1}, 2\times10^{2})$, $\rho_g=0.01$, $\rho_y=10$, and $N=20$.

\subsection{Data Generation}\label{sec:amount_of_data}

For \textit{in vivo} experiments, it is of interest to minimise the duration of data collection in prior experiments, $T_s \times T$. The minimum number of input-output samples required for implementing~\eqref{eq:deepc} is given by
\begin{align}\label{eq:datalength}
T \ge (n_u + 1)(T_\text{ini} + N + n_x) - 1,
\end{align}
where $T_\text{ini}$ must be greater than the lag $\ell$ of the system~\citep{DEEPC}. While the original model has $n_x=18$ states, Fig.~\ref{fig:fig1}.b shows that \SI{99.4}{\percent} of the dynamics are captured by the 5 most dominant modes, so $n_x=5$ is used in~\eqref{eq:datalength}. For the reduced system, it can be verified that $\ell = 5$ leads to an observability matrix that is full rank, so $T_\text{ini}=5$. This results in the minimal amount of data given by $T\ge 3(N + 10) - 1.$ For $N=20$, the minimum amount of data required is therefore $T\geq 89$. Note that without model reduction the minimum amount of data required would be at least $T\geq 167$ (\SI{27.8}{\hour} for $T_s=\SI{10}{\minute}$).

To generate the data used in~\eqref{eq:data}, 90 sample inputs were applied from a random walk starting at $u_s = A_t$ and $u_g = A_g$, then a constant input of $u_g = A_g$ and $u_s = A_t$ was applied for another 90 samples. This is done such that the system is ``reset" to a similar starting state that can be compared across controllers; however, the controller also works without these additional 90 samples.

Although basis functions are used to address input nonlinearities in~\eqref{eq:deepc} (see Section~\ref{sec:basisfunctions}), relying solely on data collected offline for construction of the Hankel matrices can lead to poor performance. Approaches for updating the data to overcome this limitation have been proposed~\citep{online_deepc_baros_2022}, but they require that the Hankel matrix has full rank at each time step. In the present system with arbitrary reference signals, there is no guarantee that the Hankel matrix maintain full rank \textit{after} updating the data. To address this, the new input-output samples are appended to $\bm{u}^d$ and $\bm{y}^d$ at each time step. Therefore, if the initial input data $\bm{u}^d$ is persistently exciting at time $t=0$, then this persistence of excitation requirement will also be satisfied at all future times. A drawback of this approach is that the computational complexity of solving~\eqref{eq:deepc} increases as $t$ increases. However, in \textit{in vivo} experiments, where the duration is not expected to exceed \SI{96}{\hour} (576 samples for $T_s\!=\!\SI{10}{\minute}$), standard computing hardware can accommodate thousands of cells~\citep{JamesMscope2025}.

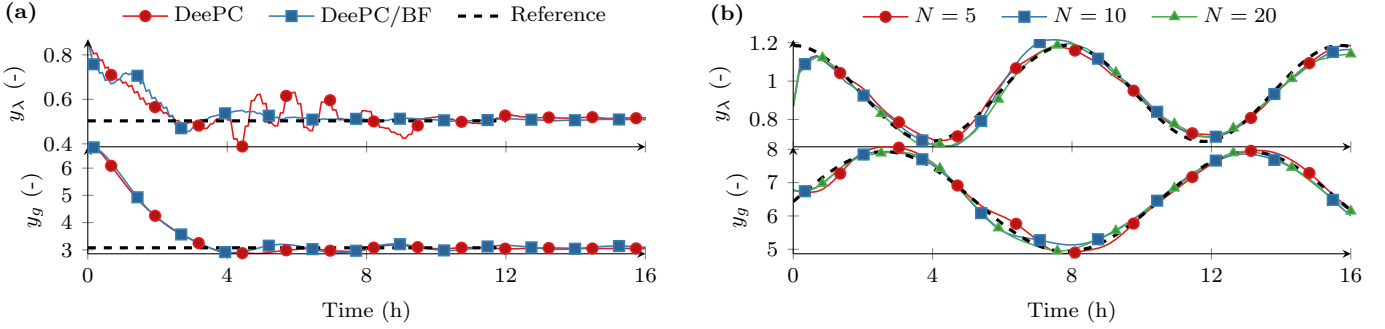
\begin{figure}%
\centering%
\begin{tikzpicture}
\begin{axis}[font=\footnotesize,at={(0, 0)},anchor=south,
    width=\columnwidth, 
    height=3cm,
    xlabel={},, xticklabels={},
    ylabel={$y_\lambda$ (-)},
    axis y line=middle, 
    axis x line=bottom,
    axis line style={-}, 
    xticklabel style={below},
    ylabel near ticks,
    xlabel near ticks,
    axis on top,
    xmin=0, xmax=16, xtick={0,4,8,12,16},
    legend style={at={(0.5,1)}, anchor=south,draw=none,font=\footnotesize,/tikz/every even column/.append style={column sep=0.25cm},inner sep=3pt},
    legend columns=3,
    enlarge y limits,
    axis x line=bottom,
    axis y line=left,
    cycle list/Set1-5,cycle multiindex* list={mark list*\nextlist Set1-5\nextlist},
]
\addplot+[thick,mark repeat=15,mark phase=9,semithick] 
table[x=t, y expr={\thisrow{yl}*100}, col sep=comma] {csv/step_no_basis.csv};
\addlegendentry{DeePC}
\addplot+[thick,mark repeat=15,mark phase=3,semithick,mark size=2] 
table[x=t, y expr={\thisrow{yl}*100}, col sep=comma] {csv/step_deepc5.csv};
\addlegendentry{DeePC/BF}
\addplot[dashed, very thick] coordinates {(0, 0.005034513642245613*100) (12, 0.005034513642245613*100)};
\addlegendentry{Reference}
\end{axis}

\begin{axis}[font=\footnotesize,at={(0, 0)},anchor=north,
    axis on top,clip=true,
    width=\columnwidth, 
    height=3cm,
    ylabel={$y_g$ (-)},
    axis y line=middle, 
    axis x line=bottom,
    axis line style={-}, 
    xticklabel style={below},
    ylabel near ticks,
    xlabel near ticks,
    xmin=0, xmax=16, xtick={0,4,8,12,16},
    enlarge y limits,
    axis x line=bottom,
    axis y line=left,
    cycle list/Set1-5,cycle multiindex* list={mark list*\nextlist Set1-5\nextlist},
]
\addplot+[thick,mark repeat=15,mark phase=9,semithick] 
table[x=t, y expr={\thisrow{yg}/10000}, col sep=comma] {csv/step_no_basis.csv};
\addplot+[thick,mark repeat=15,mark phase=3,semithick,mark size=2] 
table[x=t, y expr={\thisrow{yg}/10000}, col sep=comma] {csv/step_deepc5.csv};
\addplot[dashed, very thick] coordinates {(0, 30758/10000) (12, 30758/10000)};
\end{axis}

\node[anchor=north west, font=\footnotesize] at (current bounding box.north west) {\textbf{(a)}};
\end{tikzpicture}\\
\begin{tikzpicture}
\begin{axis}[font=\footnotesize,at={(0, 0)},anchor=south,
    width=\columnwidth, 
    height=3cm,
    xlabel={},, xticklabels={},
    ylabel={$y_\lambda$ (-)},
    axis y line=middle, 
    axis x line=bottom,
    axis line style={-}, 
    xticklabel style={below},
    ylabel near ticks,
    xlabel near ticks,
    axis on top,
    xmin=0, xmax=16, xtick={0,4,8,12,16},
    legend style={at={(0.5,1)}, anchor=south,draw=none,font=\footnotesize,/tikz/every even column/.append style={column sep=0.25cm},inner sep=3pt},
    legend columns=3,
    enlarge y limits,
    axis x line=bottom,
    axis y line=left,
    cycle list/Set1-5,cycle multiindex* list={mark list*\nextlist Set1-5\nextlist},
]
\addplot+[thick,mark repeat=10,mark phase=9,semithick] 
table[x=t, y expr={\thisrow{yl}*100}, col sep=comma] {csv/sinusoidal_deepc5.csv};
\addlegendentry{$N=5$}
\addplot+[thick,mark repeat=10,mark phase=3,semithick,mark size=2] 
table[x=t, y expr={\thisrow{yl}*100}, col sep=comma] {csv/sinusoidal_deepc10.csv};
\addlegendentry{$N=10$}
\addplot+[thick,mark repeat=10,mark phase=6,semithick] 
table[x=t, y expr={\thisrow{yl}*100}, col sep=comma] {csv/sinusoidal_deepc20.csv};
\addlegendentry{$N=20$}
\addplot[dashed, very thick]
table[x=t, y expr={\thisrow{yl}*100}, col sep=comma] {csv/sinusoidal_ref.csv};
\end{axis}

\begin{axis}[font=\footnotesize,at={(0, 0)},anchor=north,
    axis on top,clip=true,
    width=\columnwidth, 
    height=3cm,
    xlabel={Time (\si{\hour})},
    ylabel={$y_g$ (-)},
    axis y line=middle, 
    axis x line=bottom,
    axis line style={-}, 
    xticklabel style={below},
    ylabel near ticks,
    xlabel near ticks,
    xmin=0, xmax=16, xtick={0,4,8,12,16},
    enlarge y limits,
    axis x line=bottom,
    axis y line=left,
    cycle list/Set1-5,cycle multiindex* list={mark list*\nextlist Set1-5\nextlist},
]
\addplot+[thick,mark repeat=10,mark phase=9,semithick] 
table[x=t, y expr={\thisrow{yg}/10000}, col sep=comma] {csv/sinusoidal_deepc5.csv};
\addplot+[thick,mark repeat=10,mark phase=3,semithick,mark size=2] 
table[x=t, y expr={\thisrow{yg}/10000}, col sep=comma] {csv/sinusoidal_deepc10.csv};
\addplot+[thick,mark repeat=10,mark phase=6,semithick] 
table[x=t, y expr={\thisrow{yg}/10000}, col sep=comma] {csv/sinusoidal_deepc20.csv};
\addplot[dashed, very thick]
table[x=t, y expr={\thisrow{yg}/10000}, col sep=comma] {csv/sinusoidal_ref.csv};
\end{axis}

\node[anchor=north west, font=\footnotesize] at (current bounding box.north west) {\textbf{(b)}};
\end{tikzpicture}%
\caption{DeePC simulations: (a) step responses for linear DeePC and DeePC with basis functions (DeePC/BF) (both with $N=5$). (b) DeePC/BF response for $N\in\lbrace 5,10,20\rbrace$ and a sinusoidal reference.}%
\label{fig:deepc_variant_responses}%
\end{figure}%

\subsection{Basis functions for nonlinearities}\label{sec:basisfunctions}

Several variations and extensions of~\eqref{eq:deepc} have been proposed, such as online DeePC~\citep{refactoring_light_switchable_tcs_Schmidl2014}, robust DeePC~\citep{Coulson2019RegularizedAD}, or nonlinear DeePC~\citep{LI2025100219}. However, in view of future \textit{in vivo} experiments, where the controller will be applied to thousands of cells in parallel, it is of interest to keep the control algorithm both conceptually and computationally as simple as possible. One variation of nonlinear DeePC that does not complicate the controller formulation is using basis functions for input and output nonlinearities~\citep{10591192}. Here, the system is nonlinear in the inputs $u_s$ and $u_g$ (see~\appendixeqref{eqn:energy:a}{A.1a} and~\eqref{eqn:synthetic_transcription}), with the nonlinear form depending on the parameters $A_g$, $h_g$, $A_t$. Denoting the estimated parameters by $\hat{A}_{g}$, $\hat{h}_{g}$, and $\hat{A}_{t}$, we introduce the following basis functions:
\begin{subequations}\label{eq:basisfunctions}
\begin{align}
\phi_1(u_s)&\eqdef u_s / (\hat{A}_{t} + u_s),\\
\phi_2(u_g)&\eqdef (u_g/\hat{A}_{g})^{\hat{h}_{g}}/ (1+(u_g/\hat{A}_{g})^{\hat{h}_{g}}),
\end{align}
\end{subequations}
and $\phi(u)\eqdef\col(\phi_1(u_s), \phi_2(u_g))$. Note that $\phi_1$ and $\phi_2$ are monotonically increasing functions for $\hat{h}_{g}\geq 1$. With~\eqref{eq:basisfunctions}, problem~\eqref{eq:deepc} is rewritten in terms of $\bm{\delta \phi(u)}$ as
\begin{equation}\label{eq:deepc2}
\begin{aligned}
\min_{g, \bm{\delta \phi}, \bm{y},\sigma_{y}}\,& \norm{\bm{y}-\bm{r}}_{Q}^{2}+\norm{\delta\bm{\phi}}_{R^\phi}^{2} +\rho_{g}\norm{g}_{1}+\rho_{y}\norm{\sigma_{y}}_{1}\\ 
\text{s.t.}\,\,&\begin{pmatrix} U_{\text{p}}^\phi\\ Y_{\text{p}}\\ U_{\text{f}}^\phi\\ Y_{\text{f}} \end{pmatrix}g\!=\!
\begin{pmatrix} \bm{\phi}_{\text{ini}}\\ \bm{y}_{\text{ini}}\\ \bm{\phi_{t-1}}\!+\!\Delta \bm{\delta \phi}\\ \bm{y} \end{pmatrix}\!+\!\begin{pmatrix} 0\\ \sigma_{y}\\ 0\\ 0 \end{pmatrix}\!,\\
&\phi(u_\text{min})\leq \bm{\phi_{t-1}}\!+\!\Delta \bm{\delta \phi}\leq \phi(u_\text{max}),
\end{aligned}
\end{equation}
where
$\bm{\phi}\!=\!\col(\phi(u_0),\dots,\phi(u_{N-1}))$ and $\bm{\phi}\!=\!\bm{\phi_{t-1}}\!+\!\Delta \bm{\delta \phi}$, 
$R^\phi=I_N\otimes\diag(1, 10)$, $\mathcal{U}_p^\phi$ and $\mathcal{U}_f^\phi$ refer to~\eqref{eq:data} with $\phi(\cdot)$ applied element-wise to the data, and the constraints on the last line are interpreted element-wise. 

The introduction of~\eqref{eq:basisfunctions} requires some system knowledge for the form of these nonlinearities, which improves the performance of the regularised DeePC algorithm, even in the presence of uncertain parameters (see Section~\ref{sec:robustness}). Fig.~\ref{fig:deepc_variant_responses}.a compares the performance of DeePC with (DeePC/BF) and without~\eqref{eq:basisfunctions} for step reference signals, showing significantly better control for DeePC/BF. Fig.~\ref{fig:deepc_variant_responses}.b additionally compares the performance for increasing horizon lengths $N\in\lbrace 5,10,20\rbrace$ and sinusoidal reference signals. It can be seen that the control performance increases for longer horizons, which is not the case for step reference signals (not shown).

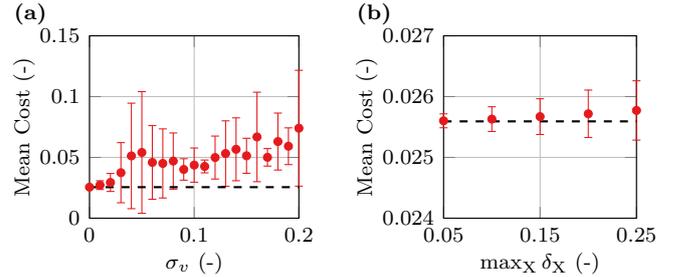
\begin{figure}
\centering
\begin{subfigure}{0.32\linewidth}%
\begin{tikzpicture}
\begin{axis}[axis on top=false,clip=false,
font=\scriptsize,
width=\linewidth,height=3.8cm,
xlabel={$\sigma_v$ (-)},
ylabel={Mean Cost (-)},
xmin=0,xmax=0.2,
ymin=0,
xtick={0,0.1,0.2},
xticklabels={0,0.1,0.2},
grid=both,
mark size=1.5pt,
xlabel style={yshift=0.25em},  
ylabel style={yshift=-0.5em},
]
\addplot[legreddraw, thick, name path=mean]
table[col sep=comma, x=noise_std,
y expr={\thisrow{mean_costs_output_only}/0.0255229999119863 - 1}
]{csv/robustness_tests_noise.csv};
\addplot[draw=none, thick, name path=upper]
table[col sep=comma, x=noise_std,
y expr={(\thisrow{mean_costs_output_only}+\thisrow{cost_std_output_only})/0.0255229999119863 - 1}
]{csv/robustness_tests_noise.csv};
\addplot[legredfill, opacity=0.2] fill between [of=upper and mean];
\end{axis}

\node[anchor=north west, font=\footnotesize,inner sep=0pt] at ([yshift=3.3cm]current bounding box.south west) {\textbf{(a)}};
\end{tikzpicture}%
\end{subfigure}
\begin{subfigure}{0.32\linewidth}%
\centering
\begin{tikzpicture}
\begin{axis}[axis on top=false,clip=false,
font=\scriptsize,
width=\linewidth,height=3.8cm,
xlabel={$\tau_g$ (\si{\minute})},
ylabel={Mean Cost (-)},
xmin=0,xmax=30,
ymin=0,
xtick={0,10,20,30},
xticklabels={0,10,20,30},
grid=both,
mark size=1.5pt,
xlabel style={yshift=0.25em},  
ylabel style={yshift=-0.5em},
]

\addplot [legreddraw,thick]
table [col sep=comma, x=tau, y expr={\thisrow{mean_cost_output_only}/0.0255229999119863 - 1}]
{csv/robustness_tests_delays.csv};
\end{axis}

\node[anchor=north west, font=\footnotesize,inner sep=0pt] at ([yshift=3.3cm]current bounding box.south west) {\textbf{(b)}};
\end{tikzpicture}%
\end{subfigure}
\hfill
\begin{subfigure}{0.32\linewidth}%
\centering
\begin{tikzpicture}
\begin{axis}[axis on top=false,clip=false,font=\scriptsize,width=\linewidth,height=3.8cm,
xlabel={$\delta_\mathrm{X}$ (\%)},
ylabel={Mean Cost (-)},
ymin=0,
xmin=5, xmax=25,
xtick={5,15,25},
xticklabels={5,15,25},
grid=both,
mark size=1.5pt,
xlabel style={yshift=0.25em},  
ylabel style={yshift=-0.5em},
]

\addplot[legreddraw, thick, name path=mean]
table[col sep=comma, x expr={\thisrow{x}*100},
y expr={(\thisrow{meanvals}-0.0256022247577895)/0.0256022247577895}
]{csv/parameter_stats.csv};
\addplot[draw=none, thick, name path=upper]
table[col sep=comma, x expr={\thisrow{x}*100},
y expr={(\thisrow{meanvals}+\thisrow{stds})/0.0256022247577895 - 1}
]{csv/parameter_stats.csv};
\addplot[legredfill, opacity=0.2] fill between [of=upper and mean];
\end{axis}
\node[anchor=north west, font=\footnotesize,inner sep=0pt] at ([yshift=3.3cm]current bounding box.south west) {\textbf{(c)}};
\end{tikzpicture}%
\end{subfigure}\\[-1em]
\caption{Mean output cost of the DeePC/BF controller and SD (shaded) for varying levels of output noise, delay, and maximal parameter uncertainties. The values are normalised and shifted by the mean cost without noise, delay, or uncertainty.}
\label{fig:robustness_results}
\end{figure}

\subsection{Robustness Analysis}\label{sec:robustness}

The present system is nonlinear and subject to measurement noise and parameter uncertainty. To evaluate the effect of measurement noise on DeePC/BF controller performance, Gaussian noise is added independently to each output. The model is evaluated on 25 step responses of different magnitudes, allowing the controller performance to be evaluated against a wide range of the expected achievable output values. A step response of $\ell_{s}=200$ samples is used, with sampling time $T_s=\SI{10}{\min}$. The cost for a given response is calculated as $c_i=\sum_{t=0}^{\ell_{s}-1} \norm{r_{i,t}-y_t}_Q^2/\ell_{s}$, $i=1,\dots,25$, and the total cost is obtained by averaging over all $c_i$. The measured output is given by $\hat{y}_t=y_t+v_t$, where $v_t\!\eqdef\!\col(v_{\lambda,t},v_{\mathrm{g},t})$ and $v_\mathrm{x}$ is distributed as $v_\mathrm{x}\sim\mathcal{N}(0, y_{\mathrm{x},t}\sigma_v)$, $\mathrm{x}\in\lbrace\lambda,\mathrm{g}\rbrace$, with $\sigma_v$ being a parameter. This distribution keeps the signal to noise ratio constant across output values. The mean cost across the 25 step responses is evaluated against $\sigma_v$ in Fig.~\ref{fig:robustness_results}.a. The experiments are repeated 20 times for each value of $\sigma_v$. The controller performance degrades as the level of noise increases; however, the performance could be enhanced by using a DeePC formulation that explicitly accommodates measurement noise~\citep{SASSELLA20231382}.

Figure~\ref{fig:robustness_results}.b investigates the impact of the delay $\tau_\mathrm{g}$ on the output cost. As this delay is not known exactly and could vary, the performance is evaluated for $\tau_\mathrm{g}\in \lbrace 0,1,\dots,30\rbrace$. The output cost increases approximately linearly with the amount of delay.

In addition to noise and delay, parameter uncertainty impacts the basis functions~\eqref{eq:basisfunctions}, which are key to obtain the constrained quadratic program~\eqref{eq:deepc2}. To investigate the impact while maintaining a similar output reachable set, we keep the same system parameters, but vary the estimated parameters $\hat{A}_g$, $\hat{h}_g$, and $\hat{A}_t$ as $\hat{\mathrm{X}} = \mathrm{X}(1 \pm\delta_\mathrm{X})$, where $\mathrm{X}$ denotes the true parameter value and $\delta_\mathrm{X}\in \lbrace 0.05,0.1,\dots,0.25\rbrace$. The incurred mean cost due to basis function parameter uncertainty is evaluated over all possible combinations of $\hat{A}_g$, $\hat{h}_g$, and $\hat{A}_t$, and shown in Fig.~\ref{fig:robustness_results}.b. The controller still reaches the desired steady-state value for these variations, although the cost tends to increase as $\delta_\mathrm{x}$ increases.

\section{BENCHMARKS}\label{sec:comparison}


Various control approaches have been used to regulate gene expression and growth rate. Here, DeePC (with basis functions) is benchmarked against proportional-integral (PI, model-free), SLMPC (model-based), and RL and Deep MPC controllers (data-based):

\subsubsection{Proportional-integral control}

PI control requires minimal system knowledge and has been used for optogenetic control of the \textit{CcaS/CcaR} system~\citep{automated_optogenetic_control_milias_argeitis_206}. Here, a separate PI controller is used for each input-output pairing $(y_g, u_g)$ and $(y_\lambda, u_s)$. Based on simulations, the PI gains are chosen as $K_{I, \mathrm{g}}=10^{-6}$, $K_{P,\mathrm{g}}=10^{-5}$, $K_{I,s}=4 \times 10^4$, $K_{P,s}=4 \times 10^3$. 

\subsubsection{Reinforcement learning}

Reinforcement learning (RL) learns a control policy through interactions with the environment~\citep{SuttonBarto}. One approach that has been used for controlling a genetic toggle switch is based on Q-learning~\citep{10178155}, where a Q-function is learned that represents the value of taking an action in a given state. More recently, actor–critic and policy-gradient RL approaches with continuous control policies have been applied to cybergenetic and metabolic control problems, including population setpoint tracking in microbial co-cultures~\citep{ESPINELRIOS202561} and robust dynamic metabolic control~\citep{ESPINELRIOS2026}.

\begin{figure}
\centering
\begin{tikzpicture}
\begin{axis}[font=\footnotesize,at={(0, 0)},anchor=south,
    width=\columnwidth, 
    height=3cm,
    xlabel={},, xticklabels={},
    ylabel={$y_\lambda$ (-)},
    axis y line=middle, 
    axis x line=bottom,
    axis line style={-}, 
    xticklabel style={below},
    ylabel near ticks,
    xlabel near ticks,xlabel style={yshift=0.2em},
    axis on top,
    xmin=0, xmax=16, xtick={0,4,8,12,16},
    legend style={at={(1.02,1.2)}, anchor=east,draw=none,font=\tiny,/tikz/every even column/.append style={column sep=0.0cm},inner sep=0pt},
    legend columns=5,
    enlarge y limits,
    axis x line=bottom,
    axis y line=left,
    cycle list/Set1-5,cycle multiindex* list={mark list*\nextlist Set1-5\nextlist},
ylabel style={yshift=-0.25em},
]
\addplot+[thick,mark repeat=10,mark phase=9] 
table[x=t, y expr={\thisrow{yl}*100}, col sep=comma] {csv/sinusoidal_controllers_deepc.csv};
\addlegendentry{DeePC/BF}
\addplot+[thick,mark repeat=10,mark phase=3,mark size=2] 
table[x expr={\thisrow{t_ddpg_1k_ch}/60}, y expr={\thisrow{lambda_ddpg_1k_ch}*100}, col sep=comma] {csv/sinusoidal_response_ddpg_1k_ch_downsampled_interp.csv};
\addlegendentry{DDPG}
\addplot+[thick,mark repeat=10,mark phase=3,mark size=2] 
table[x=t, y expr={\thisrow{yl}*100}, col sep=comma] {csv/sinusoidal_controllers_slmpc.csv};
\addlegendentry{SLMPC}
\addplot+[thick,mark repeat=10,mark phase=6,semithick] 
table[x expr={\thisrow{Time}/60}, y expr={\thisrow{lambda}*100}, col sep=comma] {csv/DeepMPC_sinusoidal_result_1.csv};
\addlegendentry{Deep MPC}
\addplot+[thick,mark repeat=10,mark phase=6,semithick,gray] 
table[x=t, y expr={\thisrow{yl}*100}, col sep=comma] {csv/sinusoidal_controllers_pi.csv};
\addlegendentry{PI}
\addplot[dashed, very thick]
table[x=t, y expr={\thisrow{yl}*100}, col sep=comma] {csv/sinusoidal_controllers_ref.csv};
\end{axis}

\begin{axis}[font=\footnotesize,at={(0, 0)},anchor=north,
    axis on top,clip=true,
    width=\columnwidth, 
    height=3cm,
    xlabel={Time (\si{\hour})},
    ylabel={$y_g$ (-)},
    axis y line=middle, 
    axis x line=bottom,
    axis line style={-}, 
    xticklabel style={below},
    ylabel near ticks,
    xlabel near ticks,
    xmin=0, xmax=16, xtick={0,4,8,12,16},
    enlarge y limits,
    axis x line=bottom,
    axis y line=left,
    cycle list/Set1-5,cycle multiindex* list={mark list*\nextlist Set1-5\nextlist},
ylabel style={yshift=-0.25em},
]
\addplot+[thick,mark repeat=10,mark phase=9,semithick] 
table[x=t, y expr={\thisrow{yg}/10000}, col sep=comma] {csv/sinusoidal_controllers_deepc.csv};
\addplot+[thick,mark repeat=10,mark phase=9,semithick] 
table[x expr={\thisrow{t_ddpg_1k_ch}/60}, y expr={\thisrow{Pg_ddpg_1k_ch}/10000}, col sep=comma] {csv/sinusoidal_response_ddpg_1k_ch_downsampled_interp.csv};
\addplot+[thick,mark repeat=10,mark phase=3,semithick,mark size=2] 
table[x=t, y expr={\thisrow{yg}/10000}, col sep=comma] {csv/sinusoidal_controllers_slmpc.csv};
\addplot+[thick,mark repeat=10,mark phase=6,semithick] 
table[x expr={\thisrow{Time}/60}, y expr={\thisrow{Pg}/10000}, col sep=comma] {csv/DeepMPC_sinusoidal_result_1.csv};
\addplot+[thick,mark repeat=10,mark phase=6,semithick,gray] 
table[x=t, y expr={\thisrow{yg}/10000}, col sep=comma] {csv/sinusoidal_controllers_pi.csv};
\addplot[dashed, very thick]
table[x=t, y expr={\thisrow{yg}/10000}, col sep=comma] {csv/sinusoidal_controllers_ref.csv};
\end{axis}

\end{tikzpicture}\\[-1em]
\caption{Comparison of different control approaches for sinusoidal references. Both DeePC and decoupled DeePC use basis functions.}
\label{fig:controller_responses}
\end{figure}
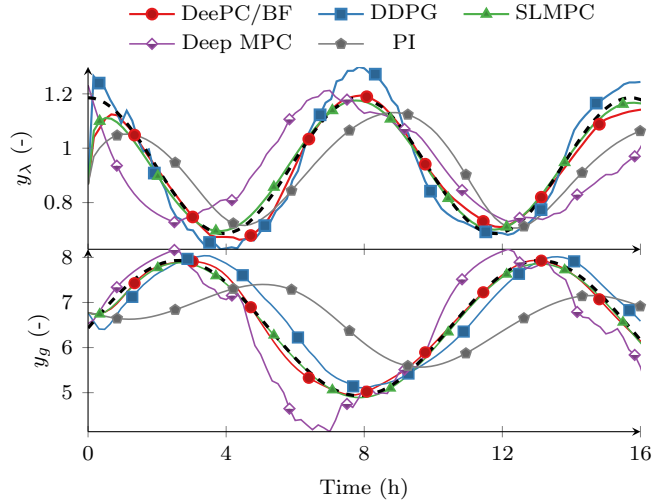

In this paper, we use the deep deterministic policy gradient (DDPG) approach~\citep{lillicrap2019continuouscontroldeepreinforcement}, an actor-critic method. The state vector is chosen to consist of the previous $10$ inputs and outputs, and the reference value. The agent is trained to track constant references, but the same training method could be used to track time-varying references such as the one from Fig. \ref{fig:controller_responses}. We evaluated control performance for up to $2,000$ episodes, and the performance plateaued after around $1000$ episodes. For the following simulations, $1000$ episodes of length $200$ were used to train the controller, with the same neural network architecture used for both actor and critic. As in~\citep{lillicrap2019continuouscontroldeepreinforcement}, experience replay is used to improve sample efficiency, and decaying action noise is added for exploration~\citep{hollenstein2023actionnoiseoffpolicydeep}.

\subsubsection{Deep Model Predictive Control}

Deep MPC is a variant of MPC that replaces the system model with a neural network (NN), and solves the resulting optimisation problem with e.g. a particle swarm optimiser. It has been applied to control the SISO \textit{CcaS/CcaR}, and we implement it in the same way for the MIMO system as described in~\citep{deep_mpc_thousands_of_single_cells_Lugagne2024}, where the inputs are assumed to take on binary values. Here, we assume that the inputs can take on the values $u_\mathrm{g}/\bar{u}_\mathrm{g} \in \lbrace 0, 4\rbrace$ and $u_s/\bar{u}_s \in \lbrace 1000, 2000\rbrace$. The NN is trained to predict both outputs based on $n_{NN}$ simulated responses of~\eqref{eq:nonlinearsys} to random input trajectories of length $\ell_{NN}=180$ ($T_s=\SI{10}{\minute}$).  We evaluated the NN's prediction accuracy for different amounts of training data ranging from $n_{NN}=1$ to $n_{NN}=2200$ input trajectories and found performance plateaued for $n_{NN}\geq 1000$. For the following simulations, the NN has therefore been trained using $n_{NN}\geq 1000$ (180,000 input-output samples).

\subsubsection{Successive Linearisation Model Predictive Control}

Successive linearisation MPC (SLMPC)~\citep{slmpc_zhakatayev2017} linearises the nonlinear system~\eqref{eq:nonlinearsys} at each time $t$ around $x_{t}$ and $u_{t-1}$, and solves a standard linear MPC problem. If the true system were linear, SLMPC would be equivalent to linear MPC, which has been shown to be equivalent for DeePC for deterministic LTI systems \citep{DEEPC}. Therefore, this method can be used to quantify the degradation in performance due to using the regularised DeePC formulation compared to an MPC approach with full-state feedback. As for DeePC/BF, basis functions are applied to the inputs before linearisation. Additionally, the change in inputs $\delta \phi(u)$ is penalised, such that the SLMPC problem can be written in a similar form to~\eqref{eq:deepc2} with $N=20$.


The different control approaches are compared in Fig.~\ref{fig:controller_responses} for sinusoidal reference signals. These results show that DeePC/BF slightly underperforms SLMPC. However, SLMPC requires full state feedback, accurate knowledge of the nonlinear system~\eqref{eq:nonlinearsys} and its parameters (which may vary from cell to cell), as well as linearising~\eqref{eq:nonlinearsys} at each time step. We would expect SLMPC to perform worse if the model and its parameters were not accurately known. Fig.~\ref{fig:controller_responses} also shows that the PI and Deep MPC controller are unable to accurately track the reference. For Deep MPC this can be attributed to the binary input values, but using continuous input values would require even more training data than the 180,000 input-output samples used here, which is already significantly more than the 180 samples used for DeePC/BF. DDPG is able to track time-varying references, despite only being trained on fixed references. The performance of all algorithms could potentially be improved after additional tuning and training.

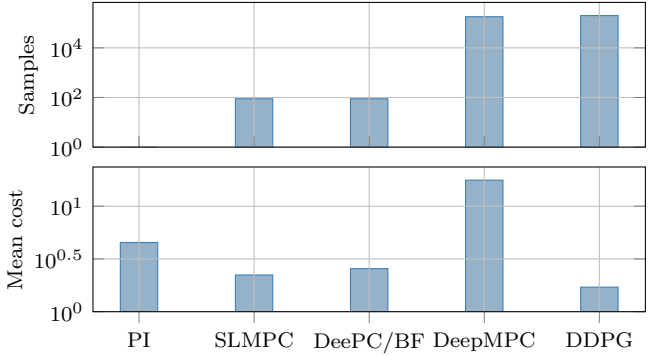
\begin{figure}
\centering
\begin{tikzpicture}
\begin{axis}[font=\footnotesize,at={(0, 0)},anchor=south east,
    ylabel style={yshift=-0.25em},
    axis on top,
    axis on top,clip=true,
    ybar,
    width=0.49\linewidth,
    height=3.25cm,
    ymin=0,
    ylabel={Samples},
    symbolic x coords={A,B,C,D,E},
    xtick=data,
    ytick={0,2,4},
    yticklabels={$10^0$,$10^2$,$10^4$,$10^{5.5}$},
    bar width=7pt,
    grid=both,
xticklabel pos=bottom,
tick align=inside,
    xtick pos=bottom,
]
\addplot+[fill=legbluefill!50, draw=legbluedraw,]
coordinates {
    (A,0)
    (B,1.949)
    (C,1.949)
    (D,5.255)
    (E,5.301)
};

\end{axis}

\begin{axis}[font=\footnotesize,at={(5em, 0)},anchor=south west,
    ylabel style={yshift=-0.25em},
    axis on top,clip=true,
    ybar,
    width=0.49\linewidth,
    height=3.25cm,
    ymin=0,
    ylabel={Mean cost},
    ytick={0,0.5,1},
    yticklabels={$10^0$,$10^{0.5}$,$10^1$},
    symbolic x coords={A,B,C,D,E},
    xtick=data,
    bar width=7pt,
    grid=both,
xticklabel pos=bottom,
tick align=inside,
    xtick pos=bottom,
]
\addplot+[fill=legbluefill!50, draw=legbluedraw,]
coordinates {
    (A,0.656)
    (B,0.348)
    (C,0.408)
    (D,1.247)
    (E,0.233)
};
\end{axis}
\node[font=\scriptsize,anchor=south east,align=left,inner sep=0] at ([yshift=0.6em]current bounding box.north east)
{
A: PI \quad
B: SLMPC \quad
C: DeePC/BF \quad
D: DeepMPC \quad
E: DDPG
};
\end{tikzpicture}
\caption{Comparison of number of input-output samples (left) required to train or design a controller and mean cost (right, scaled by $10^2$) across the 25 sample responses. It is assumed that SLMPC requires the same number of input-output samples as DeePC/BF.}
\label{fig:controller_performance}
\end{figure}

Figure~\ref{fig:controller_performance} (right) compares the control approaches for the step responses from Section~\ref{sec:robustness} in terms of mean output cost. These results show again that DeePC/BF (mean cost $2.56 \pm 2.90$ SD) and SLMPC ($2.23 \pm 2.37$) perform similarly well. For step responses, DDPG ($1.71 \pm 1.82$) performs slightly better than DeePC/BF on average, but is not able to reach all references without error. As for the sinusoidal response, the PI ($4.53 \pm 4.02$) and Deep MPC ($17.67 \!\pm\!13.27$) controllers perform significantly worse than the other controllers. As the control inputs are binarised for the Deep MPC approach, the controller is not able to reach the desired reference in all cases, so these metrics are omitted. 

Figure~\ref{fig:controller_performance} (left) compares the control approaches in terms of number of input-output samples required to implement or train the corresponding algorithm. It is assumed that SLMPC requires the same number of input-output samples, although in practice, more samples might be required to identify the parameters of the non-linear system. Defining sample efficiency as mean cost divided by number of samples required, it can be seen that for the present system, DeePC/BF is significantly more sample efficient than DDPG and Deep MPC. However, while the number of samples required by DeePC/BF varies with the controlled system (depending on $\ell$ and $n_x$), DDPG and Deep MPC might benefit from transfer learning, i.e.\ require fewer samples to be retrained on a new system.


\section{CONCLUSION}

In this paper, we have derived a model for simultaneous control of synthetic gene expression and host growth  using nutrient media concentration and optogenetic inputs. To control this two-input two-output system, linear DeePC was combined with basis functions to accommodate input nonlinearities. Additionally, model reduction was applied to reduce the amount of data for DeePC. The robustness of the DeePC controller was evaluated against delay uncertainty, basis function parameter uncertainty, and measurement noise. Future research could extend the robustness analysis to general model parameters, introduce state noise, and use the stochastic Gillespie algorithm to simulate the system dynamics.

The performance of DeePC was compared against model-free PI control, model-based SLMPC, Deep MPC, and DDPG. The simulations demonstrated that PI control is outperformed by all algorithms except for Deep MPC, which underperformed due to the use of binary inputs. The SLMPC, DeePC, and DDPG algorithms perform similarly well, but DeePC remains significantly more sample efficient than DDPG. Future research could investigate the transferability of the DDPG algorithm, i.e.\ whether the sample efficiency increases when DDPG is re-trained on a similar system.

This simulation study serves as a proof-of-concept for future \textit{in vivo} experiments on a microscopy platform capable of controlling optogenetic inputs at the single-cell level~\citep{JamesMscope2025}. In this configuration, cells are cultured in microfluidic devices while fluorescence microscopy and real-time image analysis provide measurements of gene expression and growth. Potential applications include studying cell-to-cell heterogeneity and the analysis of synthetic gene circuits under dynamically changing environmental conditions. For these experiments, minimising the data acquisition period will be critical. Future research could investigate the use of shorter sampling times $T_s$ or a sim-to-real paradigm to initialise DeePC using simulations.

\balance

\ifusebiblatex
\printbibliography
\else
\bibliography{mybibcleannodoi}
\fi

\newpage

\ifarxiv
\appendix

\section{Cell model}\label{sec:app:model}
To simulate an \textit{E.~coli} cell hosting synthetic circuitry, we use a coarse-grained resource-aware modelling framework defined based on the constraints of finite energy, ribosome, and protein pools in the cell \citep{weisse_mechanistic_links,nikolados_host_circuit_models}. 

Within the set of all considered genes $\mathcal{S}=\mathcal{S}_\text{host} \cup \mathcal{S}_\text{host} $, the set $\mathcal{S}_\text{synth}=\{g\}$ represents the set of all synthetic genes---in our case, just GFP. The remaining set $S_\text{host}$ represents the host cell's genes. The native genes are lumped by function into four classes $\mathcal{S}_\text{host}\!\eqdef\!\{\mathrm{t}, \mathrm{m}, \mathrm{q}, \mathrm{z}\}$, each treated as a single coarse grained gene. Namely, transporter proteins ($p_\mathrm{t}$) import nutrients $u_\mathrm{s}$ into the cell; metabolic proteins ($p_\mathrm{m}$) turn internal nutrients $s$ into energy storage molecules $a$; ribosomes ($p_\mathrm{z}$) synthesise all proteins; the rest of the genome encodes
housekeeping proteins ($p_\mathrm{q}$) whose repression is kept near-constant with negative autoregulation.

The  external nutrient $u_s$ affects the cell via its energy pool. It is internalized by transporter proteins as the substrate $s$, then converted into energy-bearing molecules $a$. Both these reactions follow Michaelis-Menten kinetics as per \citep{weisse_mechanistic_links, nikolados_host_circuit_models}. The energy is consumed by the cellular processes---predominantly, translation elongation by mRNA-ribosome complexes \citep{bremer1996modulation}. Both $s$ and $a$ are also diluted by cell growth. This gives~\eqref{eqn:energy}, where $\gamma=\gamma(a)$ and $\lambda=\lambda(a,\{M_\mathrm{x}\}_{\mathrm{x}\in\mathcal{S}})$ are defined in~\eqref{eqn:translation_growth} and~\eqref{eqn:growth}.
\begin{subequations}\label{eqn:energy}
\begin{align}
\dot{s} &= p_\mathrm{t} \frac{V_\mathrm{t} u_s}{A_\mathrm{t} + u_s} - p_\mathrm{m} \frac{V_\mathrm{m} s}{A_\mathrm{m} + s} - \lambda s, \label{eqn:energy:a}\\
\dot{a} &= \eta_s p_\mathrm{m} \frac{V_\mathrm{m} s}{A_\mathrm{m} + s} - \lambda a - \sum_{\mathrm{x} \in \mathcal{S}} \gamma M_\mathrm{x},
\end{align}
\end{subequations}

Each gene $\mathrm{x}\!\in\!\mathcal{S}_\text{host}$, is expressed in two steps: transcription and translation. The former is the process of reading mRNA molecules $m_\mathrm{x}$ from a gene's DNA. Transcription rates $\alpha_\mathrm{x}$ are affected by the availability of the cell's energy-bearing molecules $a$, captured by the parameters $\theta_\mathrm{x}$ in the following definitions:
\begin{subequations}\label{eqn:host_transcription}
\begin{align}
    \alpha_\mathrm{x}(a) &\eqdef \alpha_{\mathrm{x}, \text{max}} \frac{a}{\theta_\mathrm{x} + a}, \qquad \mathrm{x} \in\mathcal{S}_\text{host}\backslash\{\mathrm{q}\},\\
    \alpha_\mathrm{q}(a, p_\mathrm{q}) &\eqdef \alpha_{\mathrm{q}, \text{max}} \frac{a}{\theta_\mathrm{q} + a} \cdot \frac{1}{1 + (p_\mathrm{q}/A_\mathrm{q}) ^ {h_\mathrm{q}}},\\
    \alpha_\text{syn}(a,u_\mathrm{g}) &\eqdef
    \frac{\alpha_{\text{syn},\text{max}}a}{\theta_\text{syn}+a} \cdot 
    \frac{F_b +\left(u_\mathrm{g}(t-\tau_\mathrm{g})\right)^{h_\mathrm{g}}}{A_\mathrm{g}+\left(u_\mathrm{g}(t-\tau_\mathrm{g})\right)^{h_\mathrm{g}}}
\end{align}
\end{subequations}
where the production rate $\alpha_\mathrm{q}$ of housekeeping mRNA includes a term for negative autoregulation by $p_\mathrm{q}$ and synthetic GFP gene transcription rate is a repeat of \eqref{eqn:synthetic_transcription}. Translation depends on a finite and limited pool of the cell's free ribosomes $p_\mathrm{z}$, which bind mRNAs at (constant) rates $\kappa^+_\mathrm{x}$ to form translational complexes $M_\mathrm{x}$. These complexes can then release mRNAs either by dissociating at rates $\kappa^-_\mathrm{x}$ or completing translation to produce a protein $p_\mathrm{x}$ at rates 
\begin{align}\label{eqn:translation_growth}
    &v_\mathrm{x}(a) \eqdef \gamma(a)/n_\mathrm{x},
    &\gamma(a) \eqdef (\gamma_\text{max} a)/(K_\gamma + a),
\end{align}
where $\mathrm{x}\in\mathcal{S}$, $n_\mathrm{x}$ are protein lengths in amino acids (\si{aa}), and $\gamma(a)$ is the energy-dependent translation elongation rate in \si{aa\per\minute}, which follows Michaelis-Menten kinetics.

The cell's protein density $\rho$ (in \si{aa} per cell) is finite and constant \citep{weisse_mechanistic_links}, so the total rate of translation elongation must match the rate of protein removal. Protein degradation in \textit{E.~coli} is negligible, so proteins are predominantly removed by dilution as the cell volume grows at the rate $\lambda$ given by \eqref{eqn:growth_again}. 
\begin{equation}\label{eqn:growth_again}
\lambda(a,\{M_\mathrm{x}\}_{\mathrm{x}\in\mathcal{S}})\eqdef 
\left(\gamma(a)/\rho\right)\sum_{\mathrm{x}\in \mathcal{S}} M_\mathrm{x},
\end{equation}

Unlike proteins, mRNAs decay at non-negligible rates $\delta_\mathrm{x}$ in addition to being diluted, we establish ordinary differential equations (ODEs) for gene expression dynamics:
\begingroup
\allowdisplaybreaks
\begin{subequations}\label{eqn:host_genes}
    \begin{align}
    \dot{m}_\mathrm{x} &= \alpha_\mathrm{x} - (\lambda + \delta_\mathrm{x} + k^+_\mathrm{x} p_\mathrm{z}) m_\mathrm{x} + \left(v_\mathrm{x} + k^-_\mathrm{x}\right) M_\mathrm{x},
    \label{subeqn:mrnas}\\
    \dot{M}_\mathrm{x} &= - \left(\lambda + v_\mathrm{x} + k^-_\mathrm{x}\right)M_\mathrm{x} + k^+_\mathrm{x} p_z m_\mathrm{x},
    \label{subeqn:translational_complexes}\\
    \dot{p}_\mathrm{x} &= v_\mathrm{x} M_\mathrm{x} - \lambda p_\mathrm{x}, 
    \label{subeqn:host_proteins}\\
    \dot{p}_z &= v_\mathrm{z} M_\mathrm{z} - \lambda p_\mathrm{z} \!+\!\!\! \sum_{\mathrm{x}\in \mathcal{S}} \left(v_\mathrm{x} M_\mathrm{x} \!-\! k^+_\mathrm{x} m_\mathrm{x} p_\mathrm{z} \!+\! k^-_\mathrm{x} M_\mathrm{x}\right),
    \label{subeqn:ribosomes}\\
    \dot{p}_\mathrm{g} &= v_\mathrm{g}(a) M_\mathrm{g} - (\lambda + \mu_\mathrm{g}) p_\mathrm{g},
    \label{subeqn:nascent_gfp}\\
    \dot{P_\mathrm{g}} &= \mu_\mathrm{g} p_\mathrm{g} - \lambda P_\mathrm{g}
    \label{subeqn:maturated_gfp}
    \end{align}
\end{subequations}
\endgroup
\noindent where $\mathrm{x}\!\in\!\mathcal{S}_\text{host}$ in~\eqref{subeqn:mrnas} and~\eqref{subeqn:translational_complexes}, $\mathrm{x}\!\in\!\mathcal{S}_\text{host}\backslash\{\mathrm{z}\}$ in~\eqref{subeqn:host_proteins}, and all parameters are constant except for those defined in~\eqref{eqn:host_transcription}-- \eqref{eqn:growth_again}.

For cybergenetic control, the outputs of the systems are the cell's growth rate, $y_\lambda = \lambda$, and the concentration of maturated GFP, $y_\mathrm{g}=P_\mathrm{g}$. For simulations, the inputs and outputs are normalised by $\bar{u}_s=\SI{1e4}{molecules}$, $\bar{u}_\mathrm{g}=A_g$, $\bar{y}_\lambda=\SI{1e-2}{\per\minute}$, and $\bar{y}_\mathrm{g}=\SI{1e4}{molecules}$.

\section{Persistence of excitation}\label{sec:app:hankel}

Let $L, T \in \mathbb{Z}_{\geq 0}$ and $L \leq T$. The signal $u=\col(u_1,\dots,u_T)$, $u_i\in\R^{n_u}$, is \textit{persistently exciting of order} $L$ if the Hankel matrix
\begin{align}
\mathscr{H}_L(u)\eqdef\begin{bmatrix}
u_1 & \dots & u_{T-L+1}\\[-0.5em]
\vdots & \ddots & \vdots\\
u_L & \dots & u_T
\end{bmatrix}
\end{align}
has full row rank~\citep{DEEPC}. Additionally, consider a system of the form~\eqref{eq:linearsysdt}. The \textit{lag} of~\eqref{eq:linearsysdt} is defined by the smallest $\ell\in\mathbb{Z}_{\geq 0}$ for which \[\rank(\col(C, CA, \dots, C A^{\ell -1}))=n_x.\]

\fi  
\end{document}